\documentstyle[aps,multicol,epsf]{revtex}
\renewcommand{\narrowtext}{\begin{multicols}{2} \global\columnwidth20.5pc}
\renewcommand{\widetext}{\end{multicols} \global\columnwidth42.5pc}
\def\inseps#1#2{\def\epsfsize##1##2{#2##1} \centerline{\epsfbox{#1}}}

\def\top#1{\vskip #1\begin{picture}(290,80)(80,500)\thinlines \put(
65,500){\line( 1, 0){255}}\put(320,500){\line( 0, 1){
5}}\end{picture}}
\def\bottom#1{\vskip #1\begin{picture}(290,80)(80,500)\thinlines \put(
330,500){\line( 1, 0){255}}\put(330,500){\line( 0, -1){
5}}\end{picture}}

\begin{document}
\draft
\title{Spin-Allowed Chern-Simons Theory of
Fractional Quantum Hall States \\ for
Odd and Even Denominator Filling Factors}
\author{Tae-Hyoung Gimm, Seung-Pyo Hong, and Sung-Ho Suck Salk}
\address{Department of Physics,
Pohang University of Science and
Technology, Pohang 790-784, Korea}
\date{March 20, 1997}

\maketitle

\begin{abstract}
By allowing the spin degrees of freedom,
we present a generalized spin allowed
$U(1)\times U(1)$
Chern-Simons theory of fractional quantum Hall
effects for odd and even denominator filling factors in single layers.
This theory 
is shown to reproduce all possible odd denominator 
filling factors corresponding to 
spin-unpolarized, partially polarized, and
fully polarized
fractional quantum Hall states.
Closely following our earlier theory,
we derive the formal expressions of electromagnetic
polarization tensors and  Hall conductivity
for the spin-unpolarized and
partially polarized fractional quantum Hall states. 
Finally we report the computed spectra of 
collective excitations for both the
even and odd denominator filling factors for
which Kohn's theorem is satisfied.
\end{abstract}

\bigskip
\def\ba{\begin{array}}
\def\ea{\end{array}}
\def\bc{\begin{center}}
\def\ec{\end{center}}
\def\be{\begin{eqnarray}}
\def\ee{\end{eqnarray}}
\def\bi{\begin{itemize}}
\def\ei{\end{itemize}}
\def\bm{\left(\ba}
\def\em{\ea\right)}
\def\bn{\begin{eqnarray*}}
\def\en{\end{eqnarray*}}
\def\bt{\begin{tabular}}
\def\et{\end{tabular}}
\def\bdes{\begin{description}}
\def\edes{\end{description}}
\def\ds{\displaystyle}
\def\sss{\scriptscriptstyle}
\def\mc{\multicolumn}
\def\nc{\nonumber \\}
\def\p{{\bf p}}
\def\A{{\cal A}}
\def\vA{{\bf A}}
\def\va{{\bf a}}
\def\tA{\tilde{{\cal A}}}
\def\cE{{\cal E}}
\def\E{{\bf E}}
\def\vD{{\bf D}}
\def\vd{{\bf d}}
\def\B{{\cal B}}
\def\vB{{\bf B}}
\def\R{{\bf R}}
\def\J{{\bf J}}
\def\j{{\bf j}}
\def\x{{\bf x}}
\def\r{{\bf r}}
\def\k{{\bf k}}
\def\F{{\bf F}}
\def\Q{{\bf Q}}
\def\q{{\bf q}}
\def\H{{\cal H}}
\def\O{{\cal O}}
\def\L{{\cal L}}
\def\Z{{\cal Z}}
\def\S{{\cal S}}
\def\N{{\cal N}}
\def\l{{\bf l}}
\def\lg{\langle}
\def\rg{\rangle}
\def\la{\leftarrow}
\def\ra{\rightarrow}
\def\Ra{\Rightarrow}
\def\da{\downarrow}
\def\lra{\leftrightarrow}
\def\Ua{\uparrow}
\def\a{\alpha}
\def\b{\beta}
\def\d{\delta}
\def\e{\epsilon}
\def\f{\frac}
\def\g{\gamma}
\def\h{\hbar}
\def\k{\kappa}
\def\l{\lambda}
\def\n{\nabla}
\def\o{\omega}
\def\p{\partial}
\def\s{\sigma}
\def\t{\times}
\def\v{\varepsilon}
\def\D{\Delta}
\def\G{\Gamma}
\def\L{\Lambda}
\def\Si{\Sigma}
\def\O{\Omega}
\def\i{\infty}
\begin{multicols}{2}

\section{Introduction}
The observed fractional
quantum Hall effects (FQHE) 
encompass a large number of both even and odd denominator
filling factors 
at  low temperatures and strong magnetic fields\cite{tsg,willet,pg}.
The quantum Hall effect (QHE) at the even-denominator
filling factor\cite{willet} of $\nu = 5/2$ appeared
amidst numerous studies\cite{pg} which pay attention to  
FQHE at odd denominator filling factors.
Tilted field experiments showed that the 5/2 FQH state is 
destroyed by Zeeman coupling\cite{eisen},
indicating  that the
incompressible state of $\nu = 5/2$
is not fully polarized.
Earlier Halperin\cite{halperin} suggested that 
even denominator FQH states can exist
if one allows extra degrees
of freedom such as spin and layer index. 
Indeed the FQHE at $\nu =1/2$ was observed in double layer systems
\cite{suen}, and 
is relatively well understood\cite{lf3,hsx}.
For single layer systems
a great deal of attention\cite{lf1,hr,hlr,jain1,bj} has been paid to 
the $\nu=1/2$ state using various theoretical schemes, including
the fermion-Chern-Simons theory proposed by Lopez and Fradkin\cite{lf1}.
Haldane and Rezayi \cite{hr} showed from 
their hollow core model study that
the $\nu=1/2$ state 
as a spin-singlet state 
can be an {\sl incompressible} state.
They suggested that 
the experimentally
observed FQHE at $\nu=5/2$
could be similarly explained.
Halperin, Lee, and Read \cite{hlr} showed 
the $\nu=1/2$ state as a 
fully polarized 
{\sl compressible} state, 
by using the Chern-Simons (CS) theory based on
Jain's composite fermion (CF) picture\cite{jain1}
of the
$\nu=1/2$ state as a Fermi liquid state of spin
polarized composite fermions.
On the other hand,
Belkhir and Jain \cite{bj} suggested 
a possibility of the spin-singlet FQH state
for the  $\nu=1/2$ state at sufficiently small Zeeman energy.
However, in the present study, we pay our prime attention
to the odd denominator filling factors by allowing spin degrees of 
freedom, including some attention to the even denominator filling factors. 

Besides the spin-unpolarized FQHE state at the even
denominator filling factor of $\nu=5/2$, some odd denominator 
FQH states
tend to be either the spin-unpolarized or partially polarized states
particularly 
at relatively
weak magnetic fields \cite{clark}. 
The typical energy gap\cite{pinczuk,mellor} of
the FQH state is $\sim 12K$ or $\sim 0.08e^2/\epsilon l_0$, and
the Zeeman gap is only  $\sim$ 3K
(for instance, for $\nu=1/3$ with $B \sim 10$ T).
For the case of vanishingly small Zeeman energy,
numerical results \cite{zhang}
were in agreement with experiments\cite{clark} in that 
the FQH states at the even numerator filling factors are 
unpolarized, whereas
the Laughlin states\cite{laughlin}
of $\nu=1/(2m+1)$ with $m$ integer are 
still fully polarized and
the FQH states at odd numerator filling factors are partially polarized.
Closely following our earlier work \cite{gs},
we present a generalized spin-allowed $U(1)\times U(1)$ CS 
theory which predicts
from a single CS coupling matrix all the known
odd denominator filling factors for
unpolarized, partially polarized, and
fully polarized spin states. 
Considering that 
the $\nu=1/2$ state as a spin singlet state \cite{bj}
may have
some relevance to the experimentally observed
$\nu=5/2$ state,
we present a Chern-Simons coupling matrix
to reproduce the even denominator filling factors of $\nu=1/2$ and $5/2$.
Focusing mainly on the odd denominator filling factors
we derive 
the formal expressions of spin-allowed electromagnetic response functions,
Hall conductivity and compressibility.
In the present work we do not allow the spin-flip and
the spin-wave mode\cite{longo}.
Thus we do not consider the promotion of an electron 
from one {\sl effective} Landau level (LL) to another by permitting a spin-flip,
nor the case of the spin reversed state at the same {\sl effective} LL 
to allow a spin-wave mode.
Finally for the sake of comparison, we discuss the computed results of 
collective excitations
for the spin-unpolarized, partially
polarized, and fully polarized 
FQH states by choosing 
odd denominator filling factors.

Our main objective of the present paper
is two-fold; one for the derivation of spin-allowed states, i.e.,
the spin-unpolarized, partially polarized, and fully polarized states 
from a single Chern-Simons
coupling matrix and the other for
the revelation of differences in collective excitations between
the cases of 
the fully polarized states and
the spin-unpolarized or partially polarized states
for the odd denominator filling factors. 
\section{Spin Unpolarized, Partially Polarized
and Fully Polarized FQHE}
\label{sec:odd}
For the odd-denominator filling factors 
Jain proposed the CF wave function\cite{jain1}
\be \label{11}
\chi_{_{p/(2mp\pm 1)}}=\prod_{j<k=1}^{N_e}(z_{j}-z_{k})^{2m}
\chi_{{\pm p}}\;\;.
\ee
Here $m$ is a positive integer;
$z_{j}=x_{j}+iy_{j}$, the coordinate of the $j$th electron and
$N_e$, the total number of electrons.
$\chi_{{\pm p}}$ represents a state of integer
quantum Hall effect(IQHE) 
with an integer filling factor $p$
and $+$ (-) stands for vortex (anti-vortex)
attachment to electrons.
Considering the spin degrees of freedom but
neglecting the Zeeman coupling, 
$\chi_{{\pm p}}$ with $p=p^\uparrow+p^\downarrow$
contains a $p_{\uparrow}$ number of 
fully occupied effective LLs for spin-up electrons
and a $p_{\downarrow}$ number of 
fully occupied effective LLs for spin-down electrons.
The expression (\ref{11}) above is rewritten \cite{jain2},
\be
\label{j2}
\chi_{_{p/(2mp\pm 1)}}=\prod_{j<k=1}^{N_e}(z_{j}-z_{k})^{2m}
\chi_{{p_{\uparrow},p_{\downarrow}}}\;.
\ee
For the case of $p_{\uparrow}=
p_{\downarrow}=p/2$ with $p$ even, 
the above state represents a spin-unpolarized
(spin-singlet) state. For 
$p$ odd it is a partially or fully polarized spin state.
One notes from (\ref{j2}) that  each electron 
of up-spin or down-spin sees an even number ($2m$)
of flux quanta attached 
to other electrons.

Based on the above CF picture\cite{jain2} in (\ref{j2}),
we present a generalized spin-allowed $U(1)\times U(1)$ Chern-Simons theory of 
FQHE in order to predict all 
the known odd denominator filling factors.
For a correlated electron system 
made of both up- and down-spin electrons 
in a magnetic field $B$, we write the action
for the systems of a single layer by allowing the 
spin degrees of freedom,
\widetext
\top{-2.8cm}
\begin{eqnarray}
\label{action}
{\cal S}&=&\int  d^3 z \; \sum _{\sigma}\; \left\{ \psi^*_{\sigma}(z)
[i D_0^{\sigma} +\mu]
\psi_{\sigma}(z)-{1 \over2 m_b}|{{\bf D}^{\sigma}}\psi_{\sigma}(z)|^2 \right\} \nonumber \\
&&+ \sum _{\sigma\sigma'}\; {\alpha^{\sigma \sigma'}\over 2}
\int d^3z\;
\epsilon^{\mu \nu \lambda} {a}_{\mu}^{\sigma} \partial_{\nu}
{a}_{\lambda}^{\sigma'} -\int  d^3z\sum_\sigma \psi^*_{\sigma}(z)
\frac{g}{2}\mu_BB\sigma^z \psi_{\sigma}(z) \\
&&-{1\over 2} \int  d^3z \int d^3z'\; \sum _{\sigma ,\sigma'}\
(|\psi_{\sigma}(z)|^2-{\bar\rho}^{\sigma})
   V_{\sigma \sigma'}(|{\bf z}-{\bf z}'|)
(|\psi_{\sigma'}(z')|^2-{\bar\rho}^{\sigma'}) \nonumber
\end{eqnarray}
\bottom{-2.7cm}
\narrowtext
with
\begin{equation}
\label{dcov}
D_{\mu}^{\sigma}=\partial_{\mu}+i{e\over c}(A_{\mu}^{\sigma}-a_{\mu}^{\sigma}).
\end{equation}
Here $\sigma$ and $\sigma'$ are the spin indices and $\sigma^z=
+1$ for up-spin and $\sigma^z= -1$ for down-spin.
$\psi_{\sigma}(z)$ is a second
quantized Fermi field;
$\mu$, the chemical potential and 
${\bar \rho}_{\sigma}$, the average 
particle density with spin $\sigma$.
$A_\mu^{\sigma}$ is the electromagnetic gauge potential and 
$a_\mu^{\sigma}$, the statistical (Chern-Simons) gauge field;
$A_\mu^{\sigma}(z)=A_\mu^{-\sigma}(z)$ 
and
$a_\mu^{\sigma}(z)\neq a_\mu^{-\sigma}(z)$ by allowing
the independent density fluctuations of spin-up electrons and spin-down electrons.
The second term in (\ref{action}) is the Chern-Simons term with
the coupling matrix, $\alpha^{\sigma \sigma'}$ to be introduced below.
The Zeeman coupling is introduced 
in the third term with $\mu_B$, the Bohr magneton.
The Zeeman energy can be 
treated as a 
spin dependent chemical potential\cite{mr}, 
thus allowing the effective chemical potential,
$\mu_\sigma=\mu+g/2\mu_BB\sigma$ in (\ref{action}) above. 

The Chern-Simons action 
in (\ref{action}) causes the statistical transmutation of
each particle, by allowing an additional exchange phase as a result
of attaching flux quanta to the particle.
In the Chern-Simons term,
we introduce the following coupling matrix,
\begin{eqnarray}
\label{t2}
\alpha^{\sigma \sigma'}={\frac{e}{\phi_0\varepsilon^2}}
\left(
\begin{array}{cc}
2 m +i\varepsilon  & -2m \\
-2m & 2 m -i\varepsilon
\end{array}
\right)\;.
\end{eqnarray}
Here $\phi_0$ is the flux unit. $m$ is an integer and $\varepsilon$,
an arbitrarily small real number which is introduced to avoid 
singularity.

The spin-allowed Maxwell's equation, i.e., Euler-Lagrange equation of
motion for the statistical gauge field $a^{\sigma}_0$ is 
from (\ref{action}),
$\rho^{\sigma}(z) - \frac{1}{e}\Big \{\alpha^{\sigma\sigma}b^{\sigma}(z)
+\frac{1}{2}(\alpha^{\uparrow\downarrow}+\alpha^{\downarrow\uparrow})
b^{-\sigma}(z) \Big \}=0$,
where ${b}^{\sigma}(z)$ is the statistical `magnetic field' given by
${b}^{\sigma}(z)={\epsilon }_{ij} {\partial}_{i} a_{j}^{\sigma}(z)$.
We now use the fact $\alpha^{\uparrow\downarrow} =
\alpha^{\downarrow\uparrow}$ to write it in a simplified form
\be
\label{con}
\rho^{\sigma}(z) = \frac{1}{e}\alpha^{\sigma\sigma'}b^{\sigma'}(z)\;.
\ee
We rewrite (\ref{con}) above
\be \label{9}
{b}^{\sigma}(z) = e(\alpha^{\sigma \sigma'})^{-1}\rho^{\sigma'}(z)
=
\phi_0 \left(
\begin{array}{cc}
2 m -i\varepsilon  & 2m \\
2m & 2 m +i\varepsilon
\end{array}
\right)\;.
\ee
Now by taking the limit of $\varepsilon \rightarrow 0$ 
we explicitly result in
\be  \label{oddcon}
b^{\sigma}(z) &=& 2m\phi_0\rho^{\sigma}(z) + 2m
\phi_0\rho^{-\sigma}(z)   \nonumber \\
&=& 2m\phi_0\Big (\rho^{\sigma}(z)+\rho^{-\sigma}(z) \Big )\;.
\ee
Now we see that the expression (\ref{oddcon}) leads to
the same interpretation of
flux attachment to electrons as the CF picture theory described earlier; 
namely, 
an even number of flux quanta
$2m$ is attached to both up- and down- spin electrons.
Even allowing independent  
density fluctuations for 
spin-up and spin-down electrons, 
we find that the symmetry, 
${b}^{\sigma}(z)={b}^{-\sigma}(z)$ holds as can be readily seen 
from the relation (\ref{oddcon}) above. 
We note from (\ref{oddcon}) that various possible spin polarized states 
can arise
depending on the individual densities of 
spin-up and spin-down electrons.
They are, namely, the spin-unpolarized, partially polarized, and
fully polarized states.

After integrating out the fermion degrees of freedom in the
spin-allowed  partition function 
${\cal Z}[A_{\mu}]=\int {\cal D} \psi^* {\cal D} \psi {\cal D}
a_{\mu}^{\sigma} \; \exp \Big (i S(\psi^*,\psi,a_{\mu}^{\sigma},A_{\mu})\Big )$ and
using (\ref{con}) for
the last term of (\ref{action}),
we obtain  
the effective action
(in the natural units of $e=c=\hbar=1$)\cite{lf3},
\widetext
\top{-2.8cm}
\begin{eqnarray}
\label{effective}
{\cal S}_{\rm eff}&=& -i \sum _{\sigma} {\rm tr} \; {\ln} \left\{
i D_0^{\sigma} +\mu_{\sigma}+
{1 \over2 m_b}({\bf D}^{\sigma})^2 \right\} \nonumber \\
&&+ \sum _{\sigma\sigma'}\; {\alpha^{\sigma \sigma'}\over 2}
\int d^3z\;
\epsilon^{\mu \nu \lambda} {a}_{\mu}^{\sigma} \partial_{\nu}
{a}_{\lambda}^{\sigma'} \\
&& -{1\over 2} \int  d^3z \int d^3z' \;\Big (\alpha^{\sigma \tau}
{b}^{\tau}(z)- {\bar\rho}^{\sigma}\Big )
V_{\sigma \sigma'}(|{\bf z}-{\bf z}'|)
\Big (\alpha ^{\sigma' \tau'} {b}^{\tau'}(z')
-{\bar\rho}^{\sigma'}\Big )\;. \nonumber
\end{eqnarray}
\bottom{-2.7cm}
\narrowtext
Using 
the saddle point approximation  for the 
stationary configurations of
${\cal S}_{\rm eff}$ with respect to the small 
fluctuations of the statistical gauge field
$a_\mu^\sigma$,
$\left.\frac{\delta \S_{\rm eff}}{\delta a_{0}^{\sigma}(z)}\right|_{\bar
{a}} = 0
\hspace{0.5cm}{\rm ~and}
\hspace{0.5cm}\left.\frac{\delta \S_{\rm eff}}{\delta {\bf a}^{\sigma}
(z)}\right|_{\bar{a}} = 0$,
we readily obtain the following mean field results,
\begin{eqnarray}
\label{mf}
< \rho^{\sigma} (z)>&=& {\alpha}^{\sigma\sigma'}< {b}^{\sigma'}(z)>
\nonumber\\
< {\bf j}^{\sigma} (z)>&=& {\alpha}^{\sigma\sigma'}
< {\bf e}^{\sigma'}_{k}(z)> \\
&-&\int d^3 z' \; \alpha^{\sigma\tau}
V_{\sigma\sigma'}(z,z') \times \nonumber \\
&&[{\alpha}^{\sigma'\tau'}<{b}_{\tau'}(z')>-{\bar \rho}^{\sigma')}]\;. \nonumber
\end{eqnarray}
where ${\bf e}(z)=-\partial_t{\bf a}(z)$ is the `statistical electric field'.
Here it should be noted that our approach differs from the CS theory of 
Lopez and Fradkin \cite{lf3,lf1,lf2}, that is, in deriving 
(\ref{mf}) above we did not take 
the gauge field shift 
and avoided the condition
of the vanishing average of fluctuating electromagnetic 
field.
Since the gauge shift alters the effective action (of course, it does not
affect the partition function),
our action without the gauge shift preserves its original characteristics
(for further details we refer readers to our earlier work \cite{gs}).
The effective (or residual) magnetic field 
$B^\sigma_{\rm eff}$
is given by the difference
between the
external magnetic field $B$ and the statistical magnetic field $b$, that is,
$B_{\rm eff}^{\sigma}=B-\langle {b}^{\sigma}\rangle
=B-(\alpha^{\sigma\sigma'})^{-1} {\bar \rho}_{\sigma'}
=\hbar\omega_{\rm eff}m_{b}c/e$.
Thus the total number of 
effective magnetic flux quanta, $N^{\sigma}_{\phi_{\rm eff}}$
seen by the composite fermions of spin $\sigma$ is,
from the use of (\ref{oddcon}), 
\be 
\label{12}
N^{\sigma}_{\phi_{\rm eff}} = N_\phi-2mN^\sigma_e-2mN^{-\sigma}_e
= N_\phi-2mN_e \ee
with $N_e=N^\uparrow_e+N^{-\downarrow}_e$.
Here $N_\phi$ is the total number of magnetic flux quanta and $N_{e}^{\sigma}$,
the total number of electrons with spin $\sigma$.

The system becomes incompressible due to the presence of an energy gap
as a result of the complete filling of
an integer number $p^\sigma$ of
effective LL's by the non-interacting composite fermions of spin $\sigma$.
Obviously $N^{\sigma}_{\phi_{\rm eff}}=N^{-\sigma}_{\phi_{\rm eff}}$
from (\ref{12}).
By definition we have $\nu=N_{e}/N_\phi=(N_{e}^\uparrow
+N_{e}^\downarrow)/N_\phi=
\nu^\uparrow+\nu^\downarrow$.
Realizing that
$p^{\sigma}=\frac{N^\sigma_{e}}{N^{\sigma}_{\phi{\rm eff}}}$ 
for the effective filling factor of composite fermions with spin $\sigma$ and
$\nu^\sigma=N^\sigma_e/N_\phi$ for the filling factor of bare electrons with
spin $\sigma$,
we obtain from (\ref{12}) 
the filling factors for the electrons of spin $\sigma$, 
\be
\label{nus} \nu^{\sigma}&=& \frac{1}{\frac{1}{p^{\sigma}}
+2m+2m\frac{N_{e}^{-\sigma}}{N_{e}^{\sigma}}}\;.
\ee
Noting that $N^{\uparrow}_{\phi_{\rm eff}}=N^{\downarrow}_{\phi_{\rm eff}}$,
we get
$\frac{N_{e}^{\uparrow}}{N_{e}^{\downarrow}} = \frac{p^{\uparrow}}{p^{\downarrow}}.$ 
Thus we obtain from (\ref{nus}), 
\be \label{nuodd}
\nu =
\frac{p^\sigma+p^{-\sigma}}{1+2m(p^\sigma+p^{-\sigma})}\;.\ee
Now with some illustrations we check 
the validity of the expression (\ref{nuodd})
whether the above expression yields all
possible spin states for odd denominator filling factors.
For the unpolarized FQH states
we have
$p^\sigma=p^{-\sigma}=n$ in (\ref{nuodd}), thus, obtain the 
even numerator filling factors,
\be
\nu = \frac{2n}{1+4mn}\;.
\ee
For example, by choosing $n=m=1$, the existence of the spin-unpolarized
$\nu=2/5$ state with even numerator is predicted.
If we take $p^{-\sigma}=p^{\sigma}+1=n$ 
for the highest effective 
LL filled with spin-down electrons in (\ref{nuodd}),
we obtain the odd numerator filling factors,
\be
\label{16}
\nu=\frac{2n-1}{1+4mn-2m} \;.
\ee
By choosing $n=2$ and $m=1$, 
the partially polarized state of
$\nu=3/7$ with odd numerator is predicted.
Finally with the choice of 
$n=1$ and thus $p^{\downarrow}=1$ and $p^\uparrow=0$,
we correctly obtain from (\ref{16}) the filling factor of
\be  \nu=\frac{1}{2m+1} \ee
for the 
well-known fully polarized Laughlin states including $\nu=1/3$.
Recently Mandal and Ravishankar
suggested a doublet model for arbitrarily polarized
FQH states\cite{mr}.
However they used two different CS coupling matrices;
one for the unpolarized states and the other for
the partially polarized states.
On the other hand, in our approach it is quite gratifying to note
that 
one can extract
from the single coupling matrix in (\ref{t2})
all possible odd-denominator filling factors for all possible spin-allowed 
states.

Soon after the experimental evidence \cite{willet}
of the incompressible state for $\nu = 5/2$, there have been
attempts for its explanation \cite{hr,bj}.
Noting that the spin unpolarized FQH state of $\nu=5/2=2+1/2$ 
at relatively low magnetic fields can 
be considered as a $\nu=1/2$ state at the 2nd LL
with a sufficiently large LL spacing compared to 
correlation energy so that filled Landau levels are inert\cite{hr}. 
Belkhir and Jain\cite{bj} showed that
the $\nu=1/2$ ground state with a short range 
interaction can be 
incompressible for a wide range of repulsive interactions.
Their trial wave function is
\widetext
\top{-2.8cm}
\be 
\label{j1}
\psi_{_{1/2}}=\Big [\prod_{j<k=1}^{N}(z_{j}-z_{k})\Big ]\;
\Big [\prod_{j<k=1}^{N/2}(z_{j}-z_{k})\Big ]\;\Big [\prod_{j<k=\frac{N}{2}+1}
^{N}(z_{j}-z_{k})\Big ]\;
\chi_{_{2}}\;.
\ee
\bottom{-2.7cm}
\narrowtext
Here $z_{j}$'s with $j\leq N/2$ are the positions of spin-up electrons and
$z_{j}$'s with $j\geq N/2+1$, the positions of 
spin-down electrons. $\chi_{_{2}}$ 
represents the IQH state with 
two filled effective Landau levels, one for spin-up electrons and
the other for spin-down electrons.
$\psi_{_{1/2}}$ above is a spin-singlet state
corresponding to the fully occupied
lowest LL for the $\nu=1/2$ state.

In order to readily 
visualize the above trial state in terms of the CF picture,
we choose an electron of, say, spin-up with coordinate $z_1$. Then
$\psi_{_{1/2}}$ above contains a factor\cite{bj} 
\begin{eqnarray*}
&&(z_1-z_2)^2(z_1-z_3)^2\ldots(z_1-z_{\frac{N}{2}})^2\times \nonumber \\
&& \times(z_1-z_{\frac{N}{2}+1})(z_1-z_{\frac{N}{2}+2})\ldots(z_1-z_{N})\;\;.
\end{eqnarray*}
We are now able to easily understand from this expression that 
two kinds of flux attachments are 
`seen' by the electron of up-spin. Namely,
this electron sees an even number (two) of
flux quanta attached to each up-spin electron and 
an odd number (one) of flux quanta attached to each
down-spin electron.

We introduce the CS coupling constant,
\be
\label{theta}
\alpha_{\sigma \sigma'}=\frac{e}{(4 m - 1)\phi_0}
\left(
\begin{array}{cc}
2 m  & -(2m-1) \\
-(2m-1) & 2 m
\end{array}
\right)
\ee
where $m$ is a positive integer. 
It is easy to see that 
the CS action with $\alpha_{\sigma \sigma'}$ 
above leads to the same interpretation of flux
attachment as the composite fermion picture of
the $\psi_{1/2}$ wave function shown in the expression
(\ref{j1}).
The above spin-allowed CS coupling matrix 
is similar, in form, to the $(3,1,1)$ mode
treated for bilayer systems \cite{lf3,hsx} with the
fully polarized spin configuration.
Following a procedure similar to the case of the odd 
denominator filling factors, we obtain
for the FQHE systems of even denominator filling factors,
\be  \label{cst}
b^{\sigma}(z) &=& 2m\phi_0\rho^{\sigma}(z) + (2m-1)
\phi_0\rho^{-\sigma}(z)\;.
\ee
This result is in complete agreement with the CF 
picture of Belkhir and Jain 
shown in (\ref{j1}). That is,
each electron with spin ${\sigma}$ 
($\uparrow$ or $\downarrow$)
sees the $2m$ number of statistical flux quanta 
attached to other electrons of the same spin $\sigma$ and 
the $(2m-1)$ number of statistical flux quanta 
attached to the electrons of the opposite spin $-\sigma$. 
Allowing independent density fluctuations for spin-up and
spin-down electrons, we readily find from (\ref{cst}) that
$b^{\sigma}(z) \neq b^{-\sigma}(z)$ in general.

The total number of
effective magnetic flux quanta
seen by the composite fermions of spin $\sigma$ is given by
\be \label{neff} N^{\sigma}_{\phi_{\rm eff}} &=& N_\phi-(2m)N_{e}^\sigma
-(2m-1)N_{e}^{-\sigma} \nonumber \\
&=& N_\phi-(2m)N_{e}+N_{e}^{-\sigma}\;. \ee
The filling factor $\nu$ is  then, with 
$\nu^{\sigma}=p^{\sigma}/\Big( 1+2m p^{\sigma} + (2m-1)p^{-\sigma} \Big )$,
\be \label{nu2}
\nu &=& \nu^{\uparrow} + \nu^{\downarrow} \\ \nonumber
&=& \frac{p^\uparrow}{1+2mp^\uparrow+(2m-1)p^\downarrow} + 
\frac{p^\downarrow}{1+2mp^\downarrow+(2m-1)p^\uparrow}\;. \ee

Indeed, 
only the unpolarized states can arise, that is,
the relation (\ref{cst}) allows 
$\langle \rho^{\uparrow}(z) \rangle = \langle \rho^{\downarrow}(z)\rangle$
for the single layer constraint of
$\langle {b}^{\sigma}(z) \rangle = \langle {b}^{-\sigma}(z) \rangle$.
Quite encouragingly, we find from (\ref{nu2})
that the spin singlet (spin-unpolarized) state of $p^\sigma = p^{-\sigma}=1$
leads to the following
even denominator filling factor,
\be  \label{20}
\nu=\nu^{\uparrow}+\nu^{\downarrow}=\frac{1}{4m}+\frac{1}{4m}=\frac{1}{2m},\ee
with $\nu^{\uparrow}=\nu^{\downarrow}=1/(4m)$.
With the choice of $m=1$ the
predicted filling factor from (\ref{20}) is
$\nu=1/2$.
By setting $\nu^\sigma+1$ for the electrons of
spin $\sigma$ in the
$\nu^\sigma=1/4$ state at the second LL, we get the
$\nu=5/2$ filling factor.
It is gratifying to note that the filling factors such as
the unobserved $3/2$ and $7/2$ states can not be
predicted from the use of the CS coupling matrix in (\ref{theta}).
Unlike the case of the odd denominator filling factors for which 
all possible spin-allowed states are available with
the symmetry,
${b}^{\sigma}(z)= {b}^{-\sigma}(z)$ in the limit of 
$\varepsilon \rightarrow 0$,
the CS action with (\ref{theta}) for the even denominator filling factors is 
applicable  only
to the spin-unpolarized ground states associated 
sufficiently weak magnetic field
to preserve the symmetry condition of
$\langle {b}^{\sigma}(z) \rangle = \langle {b}^{-\sigma}(z) \rangle$ 
mentioned above.

\section{Electromagnetic Response Functions and Hall conductance}
\label{sec:hall}
We now closely follow our earlier Chern-Simons theory\cite{gs} to derive 
the electromagnetic polarization tensor, $K_{\mu\nu}$.
This approach differs from that of Lopez and Fradkin\cite{lf1} in that
our polarization tensor
represents the linear response kernel 
to the fluctuations of 
the effective gauge field rather than 
the fluctuations of the statistical gauge field
(see eq. 4.1 in Ref. 9 for comparison), as shown below. 
Allowing the Gaussian fluctuations
around the
stationary state, 
we write the spin-allowed effective action,
\widetext
\top{-2.8cm}
\begin{eqnarray}
\label{gauss}
{\cal S}_{\rm eff}({\tilde a}_{\mu}^{\sigma},{\tilde A}_{\mu}^{\sigma})&=&
      {1\over 2}  \int d^{3}x \int d^{3}y \; 
\Big ({\tilde A}_{\mu}^{\sigma}(x)-{\tilde a}_{\mu}^{\sigma}(x)\Big )\;
                  \Pi_{\mu \nu}^{\sigma\sigma'}(x,y) \;
\Big ({\tilde A}_{\nu}^{\sigma'}(y)-{\tilde a}_{\nu}^{\sigma'}(y) \Big )
\nonumber \\
&&-{1\over 2} \int  d^3x \int d^3y \;{b}_{\tau}(x)
\alpha^{\sigma\tau}
V_{\sigma \sigma'}(|{\bf x}-{\bf y}|)
\alpha^{\sigma'\tau'}{b}_{\tau'}(y) \\
&&+ {\alpha _{\sigma \sigma'}\over 2} \int d^3x\;\epsilon^{\mu \nu \lambda}
{\tilde a}_{\mu}^{\sigma}\partial_{\nu}
{\tilde a}_{\lambda}^{\sigma'}\;, \nonumber
\end{eqnarray}
\bottom{-2.7cm}
\narrowtext
The polarization tensor 
$\Pi_{\mu \nu}^{\sigma\sigma'}(x,y)$ here is simply 
the linear response kernel
to the fluctuations of the effective gauge field 
${\tilde A}^{\sigma\rm eff}_{\mu} \equiv 
{\tilde A}^{\sigma}_\mu-{\tilde a}^{\sigma}_\mu$ for the 
system made of composite fermions.
Here ${\tilde A}_{\nu}^{\sigma}$ (${\tilde a}_{\nu}^{\sigma}$)
represents the fluctuations of electromagnetic (statistical)
gauge field.
By allowing the spin degrees of freedom and using the procedure of 
Lopez and Fradkin \cite{lf1},
the Fourier transform of the polarization tensor, 
$\Pi^{\sigma\sigma'}_{\mu\nu}(q,p)=\int d^3 x d^3 y 
e^{i(q_0x_0-{\bf q}\cdot{\bf x})} e^{i(p_0y_0-{\bf p}\cdot{\bf y})}
\Pi^{\sigma\sigma'}_{\mu\nu}(x,y)$ leads to
\be \Pi^{\sigma\sigma'}_{\mu\nu}(q,p)= 
(2\pi)^3\delta^3(q+p)
\Pi_{\mu\nu}^{\sigma\sigma'}(q) \ee
with 
\widetext
\top{-2.8cm}
\be \label{free}
\Pi^{\sigma\sigma'}_{00}(q)&=&
\q^2\Pi_0^{\sigma\sigma'}(q), \nonumber \\
\Pi^{\sigma\sigma'}_{0j}(q)&=&\o q_j\Pi_0^{\sigma\sigma'}(q)
+i\epsilon_{jk}q_k\Pi_1^{\sigma\sigma'}(q), \\ 
\Pi^{\sigma\sigma'}_{j0}(q)&=&\o q_j\Pi_0^{\sigma\sigma'}(q)
-i\epsilon_{jk}q_k\Pi_1^{\sigma\sigma'}(q), \nonumber \\
\Pi^{\sigma\sigma'}_{ij}(q)&=&
\o^2\delta_{ij}\Pi^{\sigma\sigma'}_0(q)
-i\o\epsilon_{ij}q_k\Pi^{\sigma\sigma'}_1(q)+
(\q^2\delta_{ij}-q_iq_j)\Pi_2^{\sigma\sigma'}(q), \nonumber
\ee
\bottom{-2.7cm}
\narrowtext
where $\Pi_l^{\sigma\sigma'}(q)\equiv\delta_{\sigma\sigma'}
\Pi_{l;p^\sigma}(\o,\q)$
with $l=0,1,2$ and with $p^\sigma$ indicating the dependency
of the polarization tensor 
on the effective filling factor $p^\sigma$ for spin $\sigma$.

For the Gaussian integration over $a^\sigma_\mu$, we add
a gauge-fixing term
$(1/2\beta)(\partial_\mu a^\mu)^2\delta^{\sigma\sigma'}$
into (\ref{gauss}) in order to avoid singularity 
in the inversion of the matrix involving the 
quadratic term in $a^\sigma_\mu$. 
Then we obtain the following $6\times 6$ `hyper-matrix'
for the quadratic term in $a^\sigma_\mu$ in the momentum 
representation\cite{gs},
\widetext
\top{-2.8cm}
\be 
\label{66}
\tiny
M =
\left(\ba{ccc} 
\ba{c}
 \mbox{\bf q}^2\Pi_0^{\sigma\sigma'} \\ 
 + \frac{\o^2}{\beta}\delta^{\sigma\sigma'}
\ea
&
\ba{c}
 \o q_1 \Pi_0^{\sigma\sigma'} 
 - \frac{\o q_1}{\beta}\delta^{\sigma\sigma'} \\
 + iq_2 (\Pi_1^{\sigma\sigma'} + \alpha^{\sigma\sigma'}) 
\ea
& 
\ba{c}
 \o q_2 \Pi_0^{\sigma\sigma'}
 - \frac{\o q_2}{\beta}\delta^{\sigma\sigma'} \\
 -iq_1 (\Pi_1^{\sigma\sigma'} + \alpha^{\sigma\sigma'}) 
\ea
\\[15mm]
\ba{c}
 \o q_1 \Pi_0^{\sigma\sigma'} 
 - \frac{\o q_1}{\beta}\delta^{\sigma\sigma'} \\
 - i q_2 (\Pi_1^{\sigma\sigma'}+ \alpha^{\sigma\sigma'}) 
\ea
&
\ba{c}
 \o^2 \Pi_0^{\sigma\sigma'} 
 + \frac{q_1^2}{\beta}\delta^{\sigma\sigma'} \\
 + q^{2}_{2}\Big (\Pi_2^{\sigma\sigma'} - 
(\alpha^{\sigma\tau})^2 V^{\tau\sigma'}\Big ) 
\ea
&
\ba{c}
 -q_1 q_2 
 \Big (\Pi_2^{\sigma\sigma'} - (\alpha^{\sigma\tau})^2 V^{\tau\sigma'}\Big ) \\
 + \frac{q_1q_2}{\beta}\delta^{\sigma\sigma'}
 -i\o(\Pi_1^{\sigma\sigma'}+\alpha^{\sigma\sigma'}) 
\ea
\\[15mm]
\ba{c}
\o q_2 \Pi_0^{\sigma\sigma'}  
- \frac{\o q_2}{\beta}\delta^{\sigma\sigma'} \\
+ iq_1(\Pi_1^{\sigma\sigma'} +\alpha^{\sigma\sigma'}) 
\ea
&
\ba{c}
-q_1 q_2 \Big (\Pi_2^{\sigma\sigma'} -
(\alpha^{\sigma\tau})^2 V^{\tau\sigma'}\Big ) \\
+\frac{q_1q_2}{\beta}\delta^{\sigma\sigma'}
+ i \o (\Pi_1^{\sigma\sigma'}+ \alpha^{\sigma\sigma'})
\ea
&
\ba{c}
 \o^2 \Pi_0^{\sigma\sigma'} +
 \frac{q_2^2}{\beta}\delta^{\sigma\sigma'} \\
+{q}^{2}_{1}\Big (\Pi_2^{\sigma\sigma'}
- (\alpha^{\sigma\tau})^2 V^{\tau\sigma'}\Big )
\ea
\ea\right) \;.
\ee
\normalsize
\narrowtext

With $p^\uparrow=p^\downarrow$,
$\Pi^{\sigma\sigma'}_{l}$ ($l=0,1,2$) in (\ref{free}) and (\ref{66})
becomes simply the $2 \times 2$ diagonal matrix of identical elements
$\Pi^{\uparrow\uparrow}_l = \Pi^{\downarrow\downarrow}_l$
due to the involvement of the
$2\times 2$ unit matrix.
Consequently all these $2\times 2$
submatrices in
(\ref{66}) commute with each other
and this commutability aids matrix inversion.
Thus for the cases of both unpolarized ($p^\uparrow=p^\downarrow$)
and fully polarized FQH states,
the Gaussian integral over $a_\mu$ can be 
performed 
by using effectively 
reduced $3\times 3$ matrices
or by directly taking the inverse of the $6 \times 6$ matrix. 
However
for the partially polarized states
we deal directly with the above $6 \times 6$ matrix

After the Gaussian integration over 
the statistical degrees of freedom $a^\sigma_\mu$,
we obtain the effective action for the
electromagnetic fluctuations ${\tilde A}_{\mu}$,
\be \label{pole}
{\cal S}_{\rm eff}^{\rm EM} ({\tilde A}_{\mu}) 
&=&{1\over 2}\int d^{3}x \int d^{3}y \sum_{\sigma\sigma'}
{\tilde A}_{\mu}^{\sigma}(x)
                  K^{\mu \nu}_{\sigma\sigma'}(x,y)
                   {\tilde A}_{\nu}^{\sigma'} (y)\; .
\ee
Both spin-up and spin-down electrons are
coupled to the same electromagnetic fluctuations; 
${\tilde A}_{\mu}^{\uparrow}={\tilde A}_{\mu}^{\downarrow}={\tilde A}_{\mu}$.
Here $K^{\mu\nu}_{\sigma\sigma'}$ is
the electromagnetic polarization tensor
which measures the electromagnetic response of the FQHE system 
to a weak electromagnetic perturbation.
The components of the electromagnetic polarization tensor 
in the momentum space are obtained\cite{lf1} as follows:
\widetext
\top{-2.8cm}
\begin{eqnarray} \label{00}
K_{00}^{\sigma\sigma'} &=& {\q}^2  K_{0}^{\sigma\sigma'}(\omega, {\q}) \;,
\nonumber \\
K_{0j}^{\sigma\sigma'} &=& {\omega} q_{j} K_{0}^{\sigma\sigma'}(\omega, {\q})
           + i {\epsilon _{jk}}q_{k} K_{1}^{\sigma\sigma'}(\omega, {\q}) \;,
 \\
K_{j0}^{\sigma\sigma'} &=& {\omega} q_{j} K_{0}^{\sigma\sigma'}(\omega, {\q})
           - i {\epsilon _{jk}} q_{k} K_{1}^{\sigma\sigma'}(\omega, {\q}) \;,
\nonumber \\
K_{ij}^{\sigma\sigma'} &=& {\omega}^2 {\delta _{ij}} K_{0}^{\sigma\sigma'}(\omega,
  {\q}) - i {\epsilon _{ij}} {\omega} K_{1}^{\sigma\sigma'}(\omega, {\q})
+({\q}^2 {\delta _{ij}}- {q_i}{q_j})K_{2}^{\sigma\sigma'}(\omega, 
{\q})\;. \nonumber
\end{eqnarray}

For the case of 
spin-unpolarized states, we obtain for $K_{l}^{\sigma\sigma'}(\omega,\q)$ 
($l=0,1,2$)
\be  \label{kl}
K_0^{\sigma\sigma'}(\omega,\q) &=& 
- \Pi_0^{\tau\tau'}(\alpha^{\sigma\tau})^2
\Big (D^{\tau'\sigma'}(\omega,\q)\Big )^{-1}, \nonumber \\
K_1^{\sigma\sigma'}(\omega,\q)&=& 
\alpha^{\sigma\sigma'}+(\alpha^{\sigma\tau})^2
\Big (D^{\tau\tau'}(\omega,\q) \Big )^{-1}
\Big [(\alpha^{\tau'\sigma'}+\Pi_1^{\tau'\sigma'})+
\q^2(\alpha^{\sigma\tau})^3V(q)_{\tau\tau'} \Big ],
\\
K_2^{\sigma\sigma'}(\omega,\q)&=&
-(\alpha^{\sigma\tau})^2
\Big (D^{\tau\tau'}(\omega,\q)\Big )^{-1}
\Big [\Pi_2^{\tau'\sigma'}+
V(q)_{\tau'\zeta}(\omega^2
(\Pi_2^{\zeta\sigma'})^2-
(\Pi_1^{\zeta\sigma'})^2+
\q^2 \Pi_0^{\zeta\tau'}\Pi_2^{\tau'\sigma'})\Big ] \;. \nonumber
\ee
Here
$D^{\sigma\sigma'}(\omega, \q)$ 
is given by
\be \label{det}
 D^{\sigma\sigma'}(\omega, \q)=\omega^2\Pi_0^{\sigma\tau}
\Pi_0^{\tau\sigma'}-(\Pi_1^{\sigma\sigma'}+\alpha^{\sigma\sigma'})^2
+\Pi_0^{\sigma\tau}(\Pi_2^{\tau\sigma'}-\alpha^{\tau\zeta} V_{\tau\zeta}
\alpha^{\zeta\sigma'})\q^2. \ee
\bottom{-2.7cm}
\narrowtext
We obtain, in the limit of ${\q}^2 \rightarrow 0$ and
$\omega \rightarrow 0$,
\be \label{pi}
\Pi_0^{\sigma\sigma'}(0,0) &=& \frac{p^\sigma m_b}{2\pi B^{\rm eff}}
\delta^{\sigma\sigma'}=\frac{p^\sigma m_b}{2\pi (B-b)}
\delta^{\sigma\sigma'}, \nonumber \\
\Pi_1^{\sigma\sigma'}(0,0) &=& \frac{p^\sigma}{2\pi}\delta^{\sigma\sigma'},   \\
\Pi_2^{\sigma\sigma'}(0,0) &=&
-\f{(p^\sigma)^2}{2\pi m_b}\delta^{\sigma\sigma'}, \nonumber \ee
where $m_b$ is the band mass. 
Here it should be noted that the denominator in 
$\Pi_0^{\sigma\sigma'}(0,0)$ above
is given by $B^{\rm eff}=B-b$, that is,
the effective magnetic field seen by the composite fermions.
On the contrary, it is given by
the statistical magnetic field in  the CS theory of Lopez and Fradkin 
(see their equation, B3 in Ref. 9).

Using (\ref{t2}), (\ref{det}) and (\ref{pi}),
we obtain from (\ref{kl}) with $p^\uparrow = p^{\downarrow} = n$,
\widetext
\top{-2.8cm}
\be 
\label{k11}
\lim_{\o \rightarrow 0}\lim_{\q\rightarrow 0}K_1^{\sigma\sigma'}
(\omega,{\bf q})
&=&
\alpha^{\sigma\tau}\Big (\alpha^{\tau\tau'}+\Pi_1^{\tau\tau'}(0,0)\Big )^{-1}
\Pi_1^{\tau'\sigma'}(0,0)\\ \nonumber 
&=& \frac{1}{2\pi}\frac{1}{1+2m(2n)+\varepsilon^2n^2}
\left( \begin{array}{cc}
n+2mn^2
+i\varepsilon n^2   & -2mn^2 \\
-2mn^2 &
n+2mn^2
-i\varepsilon n^2
\end{array}    \right) \;, \ee
\bottom{-2.7cm}
\narrowtext
for the odd-denominator filling factors.
We obtain 
in the limit of 
${\bf q}^2 \rightarrow 0$ and $\omega\rightarrow 0$,
\be
  K_1(0,0) &=& \sum_{\sigma\sigma'}K_1^{\sigma\sigma'}(0,0)=
\lim_{\varepsilon\rightarrow 0}\frac{1}{2\pi}
\frac{2n}{1+2m(2n)+\varepsilon^2n^2 } \\ \nonumber
&=& \frac{1}{2\pi}\frac{2n}{1+4mn}
={\nu\over 2 \pi}\;. \ee
For a weak external electric field ${\tilde E}_j$, the
induced current is
$J_i=  K_1(0,0) \epsilon _{ji} {\tilde E}_j $.
Thus $ K_1(0,0)$ is the same as the 
Hall conductance $\sigma _{xy}$ of the FQHE system,
which correctly represents a {\sl fractional} multiple of ${e^2\over h}$. 
Using (\ref{kl}), (\ref{det}) and (\ref{pi}) for (\ref{00}) 
the compressibility for the unpolarized states is simply
$ \kappa = \lim_{q\rightarrow 0}K_{00}(\omega,\q)=0\;$, indicating
the incompressible states.

Likewise for the even-denominator filling factors,
\widetext
\top{-2.8cm}
\be
\label{k12}
\lim_{\o \rightarrow 0}\lim_{{\bf q}\rightarrow 0}K_1^{\sigma\sigma'}
(\omega,{\bf q})
&=& \frac{1}{2\pi}\frac{1}{1+2m(2n)+(4m-1)n^2}
\left( \begin{array}{cc}
n+2mn^2 & 
-(2m-1)n^2 \\
-(2m-1)n^2 & n +2mn^2
\end{array}    \right)\;, \ee 
\bottom{-2.7cm}
\narrowtext
and $K_1(0,0)=1/(2\pi2m)$ in this case with $n=1$.
Although 
we do not write the explicit 
form of $K_{l}^{\sigma\sigma'}(\omega,\q)$ 
($l=0,1,2$) here 
for the partially polarized states,
one can find from a similar procedure 
the Hall conductance and incompressibility.

\section{computed spectra of collective excitations}
\label{sec:coe}
There have been several theoretical attempts to obtain the
collective excitations of fully polarized states.
By using the single mode approximation (SMA)
developed in analogy with the Feynman's theory of superfluid helium,
Girvin, MacDonald, and Platzman \cite{gmp} calculated
collective excitations  of  the Laughlin ($\nu = 1/(2m+1)$) states
with $m$ positive integer and found the existence 
of relatively deep minima in the energy dispersion for $m=3, 5$ and $7$.
Recently Kamilla, Wu, and Jain used the CF wave functions to
investigate the collective excitations of various FQH states \cite{kamilla}
and also found the deep minima.
Based on the CF picture,
Chern-Simons field theoretical approaches have been used to obtain the
collective excitations of the FQH states;
Lopez and Fradkin found a series of collective modes for
fully polarized incompressible states \cite{lf2}, and
taking into account large mass renormalization,
Simon and Halperin 
\cite{sh} modified
the earlier random phase approximation (RPA)
of Halperin, Lee, and Read \cite{hlr}.
They obtained 
a correct frequency scale for low-energy excitations and
found a shallower magnetoroton minimum for the $\nu=1/3$ 
Laughlin state compared to other studies \cite{gmp,kamilla}.

\subsection{Collective excitations for odd denominator FQHE}
In the odd denominator FQH states
for which the effective filling factor is an integer in the IQH
state of CF,
there exist three different cases of spin-allowed states; 
the spin unpolarized state, 
$p^{\uparrow}= p^{\downarrow}$;
the partially polarized state,
$p^{\uparrow}\neq p^{\downarrow} \neq 0$
and the fully polarized state,
$p^{\downarrow} \neq 0$ and $p^{\uparrow} = 0$.

Earlier the collective modes of 
the fully polarized FQHE were examined 
by Lopez and Fradkin \cite{lf2}. Their computed
energy dispersion was accurate only at low values of $q$
due to the use of the lowest order term in $q$.
In the present study, in order to examine collective excitations for
the wider range of $q$ 
we include higher order terms in $q$ 
and
compute the poles of $K_{00}^{\sigma\sigma'}$ numerically.
We
adopt the components of one-particle polarization tensor 
derived by Lopez and Fradkin, which can
be extended  to the following generalized expressions to allow for
the spin-unpolarized, partially
polarized, and fully polarized FQH states \cite{lzhang}, 
\widetext
\top{-2.8cm}
\be \label{aa}
\Pi_0^{\sigma\sigma'}(\o,\q) &=&
-\delta_{\sigma\sigma'}\frac{B_{\rm eff}}{(2\pi)m_b}
e^{-\bar{q}^2}\sum^{\infty}_{m=p^\sigma}\sum^{p^\sigma-1}_{m'=0}
\frac{(m-m')}{\o^2-(\o_m-\o_{m'})^2}\frac{m'!}{m!}
{\bar q}^{2(m-m')} 
[L_{m'}^{m-m'}(\bar{q}^2)]^2 \;, \nonumber \\
\Pi_1^{\sigma\sigma'}(\o,\q) &=&
-\delta_{\sigma\sigma'}\frac{B^2_{\rm eff}}{(2\pi)m_b^2}
e^{-\bar{q}^2}\sum^{\infty}_{m=p^\sigma}\sum^{p^\sigma-1}_{m'=0}
\frac{(m-m'-1)}{\o^2-(\o_m-\o_{m'})^2}\frac{m'!}{m!}
{\bar q}^{2(m-m')} 
L_{m'}^{m-m'}(\bar{q}^2) \nonumber \\
& &\times\Big \{\bar{q}^2[L_{m'}^{m-m'}(\bar{q}^2)
+2L_{m'-1}^{m-m'-1}(\bar{q}^2)(1-\delta_{m',0})]
-(m-m')L_{m'}^{m-m'}(\bar{q}^2)\Big \} \;, \\
\Pi_2^{\sigma\sigma'}(\o,\q) &=&
-\delta_{\sigma\sigma'}\frac{B^2_{\rm eff}}{(2\pi)2m_b^3}
e^{-\bar{q}^2}\sum^{\infty}_{m=p^\sigma}\sum^{p^\sigma-1}_{m'=0}
\frac{(m-m')}{\o^2-(\o_m-\o_{m'})^2}\frac{m'!}{m!}
{\bar q}^{2(m-m'-1)}\nonumber \\
& & \times[L_{m'}^{m-m'}(\bar{q}^2)+
2L_{m'-1}^{m-m'-1}(\bar{q}^2)(1-\delta_{m',0})] \nonumber \\
& & \times\Big \{\bar{q}^2[L_{m'}^{m-m'}(\bar{q}^2)
+2L_{m'-1}^{m-m'-1}(\bar{q}^2)(1-\delta_{m',0})]
-2(m-m')L_{m'}^{m-m'}(\bar{q}^2)\Big \} \nonumber
\ee
\bottom{-2.7cm}
\narrowtext

with $\bar{q}^2\equiv \q^2/(2B_{\rm eff})=\q^2/2(B-b)$.
Differences in the effective filling factors
$p^\sigma$ 
in the 
expression (\ref{aa}) here permit three different
spin allowed FQH states mentioned above.
Due to the neglect of spin wave mode and spin flip, the Zeeman coupling
dose not affect the density
response kernel since 
the spin dependent chemical
potential contribution disappears in the
denominator of $\Pi_l^{\sigma\sigma'}(\o,\q)$ ($l=0,1,2$).
In the present study, by keeping virtual excitations up to the $9$th
 effective LL in $\Pi_l(\o, q)$ in (\ref{aa}) above
 we compute collective excitations
 from the poles that appear in
 the density-density correlation functions
 $K_{00}^{\sigma\sigma'}(\omega, \q)$.

\subsubsection{Fully polarized FQH state: $p^\uparrow=0, p^\downarrow\neq 0$}
For the fully polarized Laughlin states, 
$p^\uparrow=0$, the one-particle polarization tensor (\ref{aa}) has
non-vanishing components only for spin-down electrons,
$\Pi_l^{\downarrow\downarrow}(\o,\q)$.
The poles of the density-density correlation functions
$K_{00}^{\sigma\sigma'}(\omega, \q)$ in (\ref{00}) are
obtained from the zeroes of the determinant \cite{trick} of
$[(\alpha^{\sigma \sigma'})^{-1}]^2
D^{\sigma\sigma'}(\omega,\q)$
as can be seen from (\ref{kl}) and (\ref{det}).
Obviously the expression of frequency dispersion
at small $\q$ values
for the fully polarized states
is the same as that of
Lopez and Fradkin\cite{lf2}.
Thus we avoid repetition here.
\begin{figure}
\inseps{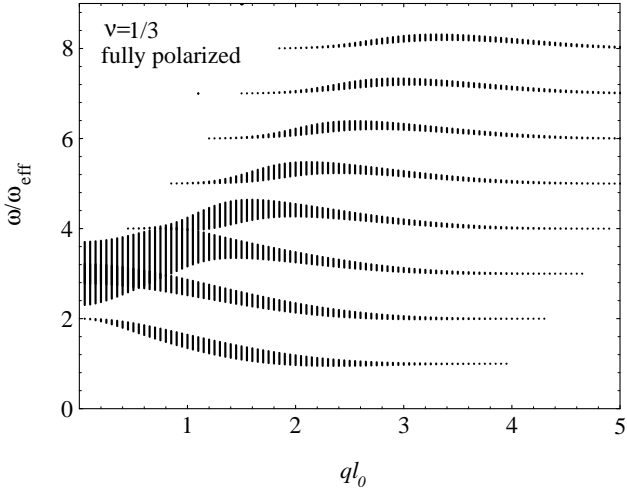}{0.5}
\caption{\label{fig1}
Collective excitation spectrum for fully polarized FQH states at
$\nu=1/3$ with the short range potential $v(0)=4/5\hbar\omega_{\rm c}$.
The center of each stripe shows
the pole and width of striped band represents
$q^{-2}$ times the weights (residues) of the pole in $K_{00}$.}
\end{figure}
In Fig.~\ref{fig1} we display
numerical calculations of collective excitation spectrum for 
$\nu=1/3$. 
As long as the short range interaction energy 
is in the order of the mean effective field gap $\omega_{\rm eff}$, we find
that the
dispersion $\o(q)$ does not change appreciably.
We choose a constant value of $v(q)=4/5\hbar\omega_{\rm c}$ with $B\sim 10$T
which is in the same order of magnitude as the Coulomb
repulsion energy $e^2/\epsilon l_0$.
Another choice of the interaction potential, 
$v(q)\sim 1/q$ does not appreciably alter the energy dispersion.
This is, indeed,
consistent with the earlier studies 
based on the Laughlin wave function\cite{laughlin}.
The computed results are in excellent agreement with 
the unrenormalized RPA results of
Simon and Halperin \cite{sh}.
In Fig.~\ref{fig1} the width of striped band represents 
the $q^{-2}$ times the weight of poles of the 
density response function $K_{00}(\o, q)$.
The width becomes negligibly small when $q$ becomes larger than $5/l_0$.
Such damped collective excitation is
already pointed out by Lopez and Fradkin\cite{lf2}, which is
thought to occur when the excitation 
energy becomes equal to an energy to create
the lowest available two-particle (particle-hole pair) state 
at sufficiently large $q$.

\subsubsection{Unpolarized FQH state: $p^\uparrow=p^\downarrow$}
To the best of our knowledge, there exists no report on comparative studies
between the spin unpolarized states 
and the spin fully polarized states
for the choice of the same filling factors.
We first 
consider the excitation mode 
for the unpolarized case with weak Zeeman coupling.
As an example, we consider
the filling factor of $\nu=2/5$
with $p^\uparrow=p^\downarrow=1$ and $m=1$ in (\ref{t2}).
The poles of the density-density correlation functions
$K_{00}^{\sigma\sigma'}(\omega, \q)$ in (\ref{00}) are
obtained from the zeroes of the determinant\cite{trick} of
$\Big [(\alpha^{\sigma \sigma'})^{-1} \Big ]^2
D^{\sigma\sigma'}(\omega,\q)$
as can be seen from (\ref{kl}) and (\ref{det})
or 
from the zeroes of the determinant of 
$6\times 6$ hyper-matrix (\ref{66})
for the spin partially polarized states.
We obtain the collective excitations 
for the unpolarized state
by finding the zeroes of
the determinant through either one of the approaches above.
\be
\det M = -\frac{(2\pi)^2(4m)^2(\o^2 + {\bf q}^2)^4}{\varepsilon^4 \beta^2}
U(\o, {\bf q})
\ee
where
\widetext
\top{-2.8cm}
\be
\label{52}
\lim_{\varepsilon\rightarrow 0}U(\o, {\bf q})&=&
\Big [\omega^2\Pi_0^2
-(\Pi_1+\frac{1}{2\pi}\frac{1}{4m})^2
+\Pi_0\Big (\Pi_2-\frac{1}{4\pi^2}\frac{1}{(4m)^2}
(v_{\uparrow\uparrow}+v_{\uparrow\downarrow})\Big )\q^2\Big ] \nonumber \\
&& \times \Big [1+\Pi_0(v_{\uparrow\uparrow}-v_{\uparrow\downarrow})\q^2 \Big ]\;, 
\ee
\bottom{-2.7cm}
\narrowtext
with $\Pi_l=\Pi_l^{\uparrow\uparrow}=\Pi_l^{\downarrow\downarrow}$.
\begin{figure}
\inseps{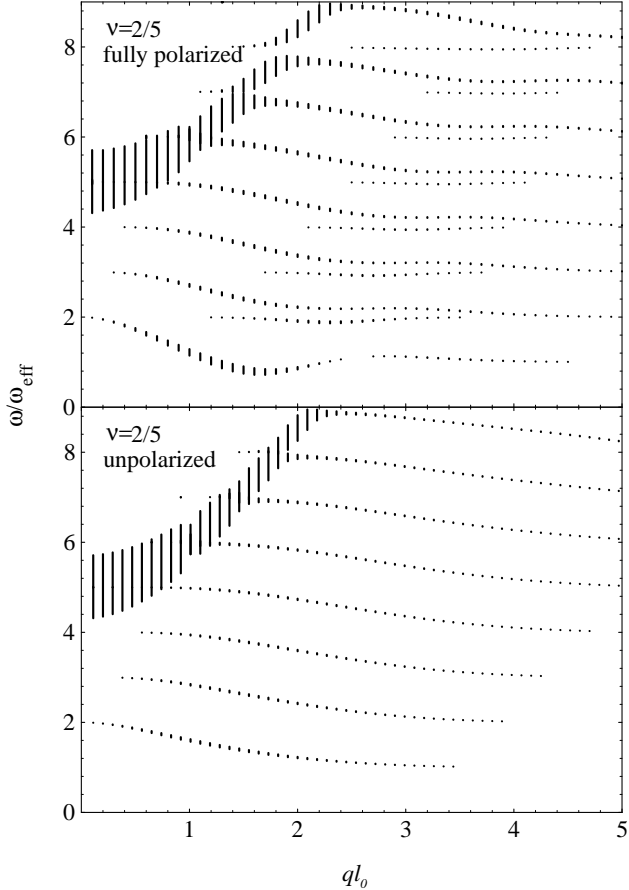}{0.5}
\caption{\label{fig2}
Collective excitation spectrum for both
the spin fully polarized and unpolarized FQH states
at $\nu=2/5$ with the short range potential $v(0)=4/5\hbar\omega_{\rm c}$.
The center of each stripe shows
the pole and width of striped band represents
$q^{-2}$ times the weights (residues) of the pole in $K_{00}$.}
\end{figure}
Here $v_{\uparrow\uparrow}$ and $v_{\uparrow\downarrow}$ 
are the zeroth order coefficients of the
Fourier transform of the interaction potential between
the composite fermions of the same spin or
opposite spins and are identical, i.e.\, 
$v_{\uparrow\uparrow}=v_{\uparrow\downarrow}=v$.
The energy dispersion relation obtained from the zeroes of (\ref{52})
then becomes similar, in form,
to the fully polarized case\cite{lf2} of $\nu=1/5$
except the difference in the potential energy dependent term
($2v$ for $\nu=2/5$ and $1v$ for $\nu=1/5$).
There exists 
only one mode whose residue is proportional to $\q^2$
at the cyclotron frequency, $\o_{\rm c}=5\o_{\rm eff}$.
By keeping only the
lowest order term in $\q$ and using (\ref{aa}) and (\ref{52}),
its frequency dispersion of $\o(q)$ 
is given by 
\be 
\label{sleep}
\omega ({\q})= \Big [ { \omega}_{c}^2 +
                          ({{\q}^2\over 2 {B_{\rm eff}}})\;
{\omega_{\rm eff}}^2(\frac{16}{3}
 +\frac{4m_bv}{2\pi})
   \Big ] ^{1\over2}\;,
\ee
with the residue (weight of poles) 
for the density response function $K_{00}$
\be
Res(K_{00}^{\sigma\sigma'},\omega _{\pm}({\q})) = -{\q}^2 \;
                                   {\omega}_{c} {\nu \over 2\pi}\;
            \left( \begin{array}{cc} 1& \;  1\\ 1& \; 1
                   \end{array} \; \right)\;,
\ee
thus clearly satisfying the Kohn's theorem\cite{kohn},
which demands that there should be only one mode converging to
the cyclotron frequency with
residue proportional to ${\bf q}^2$ in the long-wavelength limit\cite{sczhang}.
From 
the direct analogy to the fully polarized case, one
can also show that
the $\nu=2/5$ state satisfies the $f$-sum rule \cite{lf2}.

We computed 
the case of fully polarized FQH states at $\nu=2/5$ 
(see Fig.~\ref{fig2}) and found a  
clear minimum for the lowest excitation mode.
Unlike the case of
the spin fully polarized state for $\nu=2/5$ there
is an almost flat minimum for the lowest mode for the choice of 
spin unpolarized state,
as shown in Fig.~\ref{fig2}. 
One of the two modes
at the zero
momentum excitation energy of $\omega_{\rm c}=5\omega_{\rm eff}$ 
rapidly increases with  $q$;
this computed mode follows
analytic behavior of $\omega(\q)\sim \q^2$ near $q=0$.
Besides we note that there exist differences in the 
number of subbands for higher lying
collective excitation modes between the two different states.
Both cases show Kohn's modes at the long-wavelength limit.
On the other hand 
the weight of the lowest excitation mode for the fully polarized case
is somewhat larger than that of the unpolarized case.

\subsubsection{Partially polarized FQH state: $p^{\downarrow}=p^\uparrow +1$}
It is not clear\cite{wu} whether the FQHE occurs 
with the fully or partially polarized states of odd denominator 
filling factors if
the spin wave mode or spin flip excitations can happen.
However it is thought that the incompressible states of IQHE at 
odd integer filling factors are
possible if there exist large correlation effects between electrons;
Sondhi {\it et al.} \cite{sondi} showed that 
the fully polarized odd integer $\nu=1$ state
always has a gap due to the correlation effects, even if
the Zeeman energy vanishes and spin flip is allowed. 
Similarly such FQH state corresponding 
to the odd integer 
effective filling factors (e.g. $\nu=1/3$ or $3/7$)  in the CF picture
may be also possible
if there exist large correlation effects 
between composite fermions\cite{wu}. 
In the present calculations 
we did not consider 
the correlation effects between the composite fermions for
the fully or partially polarized 
FQH states.
Further we would like to point out that 
the computed energy dispersions
represent collective excitations without the consideration of
spin wave mode and spin-flip processes. 
For sufficiently large effective LL spacing,
FQH states for the odd effective filling factors 
naturally occur.

With $p^{-\sigma}=p^{\sigma}+1$ for the partially polarized states,
the diagonal elements of $\Pi^{\sigma\sigma'}_{l}$ are
no longer identical, that is, $\Pi^{\uparrow\uparrow}_{l}\neq
\Pi^{\downarrow\downarrow}_{l}$ in (\ref{aa}) due to the 
difference in population between
the up-spin electrons and down-spin electrons.
The poles of the density-density correlation functions
$K_{00}^{\sigma\sigma'}(\omega, \q)$ are
obtained from the zeroes of the determinant of 
the $6\times 6$ `hyper-matrix' (\ref{66}).
We obtain the determinant through the
symbolic calculation using {\it Mathematica}
\cite{math}, 
\be
\det M = -\frac{(2\pi)^2(2m)^2(\o^2 + {\bf q}^2)^4}{\varepsilon^4 \beta^2}
P^{\sigma\sigma'}(\o, {\bf q})
\ee
where
\widetext
\top{-2.8cm}
\be
\label{73}
\lim_{\varepsilon\rightarrow 0}P^{\sigma\sigma'}(\o, {\bf q})&=&
\omega^2(\Pi_0^\uparrow+\Pi_0^\downarrow)^2
-\Big (\Pi_1^\uparrow+\Pi_1^\downarrow
+\frac{1}{2\pi}\frac{1}{2m}\Big )^2  \nonumber \\
&&+\Big [\Pi_0^\uparrow\Pi_2^\uparrow
+\Pi_0^\uparrow\Pi_2^\downarrow+\Pi_0^\downarrow\Pi_2^\uparrow
+\Pi_0^\downarrow\Pi_2^\downarrow
-\frac{1}{4\pi^2}\frac{1}{4m^2}
(\Pi_0^\uparrow+\Pi_0^\downarrow)v^{\uparrow\uparrow}\Big ]\q^2  \\ 
&& -(v^{\uparrow\uparrow}-v^{\uparrow\downarrow})\q^2
\times\Big [\o^2\Pi^{\uparrow}_0\Pi_0^{\downarrow}
(\Pi^{\uparrow}_0+\Pi_0^{\downarrow})
-\Pi^{\uparrow}_0\Pi_0^{\downarrow}
(\Pi^{\uparrow}_2+\Pi_2^{\downarrow})\q^2  \nonumber \\
&& -\Pi_0^{\uparrow}\Pi_1^{\downarrow}\Big (\Pi_0^{\downarrow} 
+\frac{1}{2\pi}\frac{1}{2m}\Big )
-\Pi_0^{\downarrow}\Pi_1^{\uparrow}\Big (\Pi_0^{\uparrow}
+\frac{1}{2\pi}\frac{1}{2m}\Big )
-\frac{1}{4\pi^2}\frac{1}{4m^2}\Pi_0^{\uparrow}\Pi_0^{\downarrow}
(v^{\uparrow\uparrow}+v^{\uparrow\downarrow}) \Big ]  \nonumber\;.
\ee
\bottom{-2.7cm}
\narrowtext
\begin{figure}
\inseps{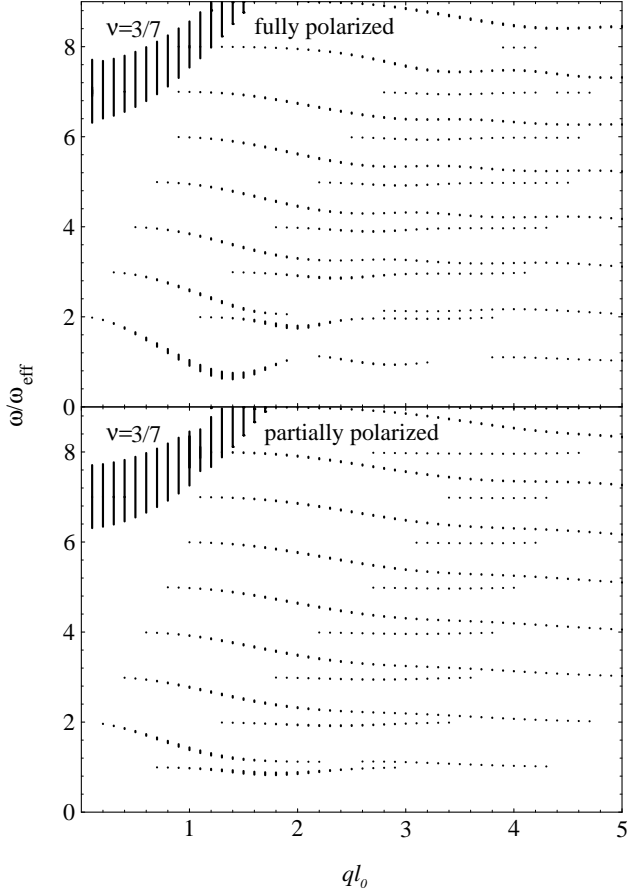}{0.5}
\caption{\label{fig3}
Collective excitation spectrum for both
the spin fully polarized and partially polarized states
at $\nu=3/7$ with the short range potential $v(0)=4/5\hbar\omega_{\rm c}$.
The center of each stripe shows
the pole and width of striped band represents
$q^{-2}$ times the weights (residues) of the pole in $K_{00}$.}
\end{figure}
The dispersion of the collective excitations 
for the $\nu=3/7$ with $m=1$ 
and $v_{\uparrow\uparrow}=v_{\uparrow\downarrow}$ is plotted in Fig.~\ref{fig3}.
The numerically obtained Kohn's mode 
$\omega=7\omega_{\rm eff}\equiv\omega_{\rm c}$
at the zero momentum is 
quite satisfactory 
with a maximum width 
as ${\q}^2\rightarrow 0$, as shown in Fig.~\ref{fig3}.
We note that
besides the differences in the energy spectrum
the lowest excitation mode 
for the spin partially polarized state of $\nu=3/7$
has a much shallower minimum compared to the case of the fully polarized state.

There exists 
discrepancy between our predicted energy gap
and the experimentally observed energy gap; 
by choosing a typical value of magnetic field $B\sim 10$T,
our predicted energy gaps 
are about $0.8\omega_{\rm eff}\sim 50$K
with band mass $m_{b}=0.066m_{e}$ ($m_{e}$, the electron bare mass)
as shown in Figs.~\ref{fig1} through \ref{fig3}
which is
about 2 to 4 times larger than the experimental values\cite{pinczuk,mellor} of
$\sim 0.1e^2/\epsilon l_0\sim 16$K
with the choice of $\epsilon=13$ for dielectric constant.

For the fully polarized states
this problem is remedied by the mass-renormalized RPA study 
of Simon and Halperin\cite{sh}.
Their finding is that the magnetoroton minima are much less pronounced
 than those calculated from the unrenormalized RPA,
and further comparison of the mass-renormalized RPA method
with the results from the 
exact diagonalization method\cite{he} showed a good agreement
particularly at low energies.
 In the present study,
 we noted marked 
 differences 
particularly in the lowest collective excitation mode between  
the fully polarized states and
the spin unpolarized or partially polarized states.
Judging from 
the shallower magnetoroton minima 
with other than the spin fully polarized states,
we believe  that a study of finite-size system  through an
exact diagonalization procedure\cite{fano} 
for the spin-unpolarized or partially polarized states 
of odd denominator filling factors may still lead to shallower magnetoroton
minima than the fully polarized case.

\subsection{Collective excitation for $\nu=1/2$ state}
For the $\nu=1/2$ state,
we consider the case of 
the effective filling factor, $p^{\uparrow}=p^{\downarrow}=1$ and $m=1$ 
in (\ref{theta}) with
sufficiently small Zeeman coupling.
The effective cyclotron frequency is 
$\omega^{\uparrow}_{\rm eff}=
\omega^{\downarrow}_{\rm eff}=\omega_{\rm c}/4$.

The poles of $K_{00}^{\sigma\sigma'}$ are obtained from
the zeroes of the determinant of (\ref{66}),
\be
\det M = -\frac{(2\pi)^2 9(\o^2 + {\bf q}^2)^4}{\beta^2}
I(\o, {\bf q})\times O(\o, {\bf q}),
\ee
where
the determinant is factorized into two parts, in order to obtain,
\be 
\label{1}
I(\o, {\bf q})&=&\omega^2\Pi_0^2
-(\Pi_1+\frac{1}{2\pi}\frac{1}{3})^2 \nonumber \\
&&+\Pi_0\Big [\Pi_2-\frac{1}{4\pi^2}\frac{1}{9}
(v_{\uparrow\uparrow}+v_{\uparrow\downarrow})\Big ]\q^2 = 0, \ee
which leads to the `in-phase' residues
and 
\be 
\label{2}
O(\o, {\bf q})&=&\omega^2\Pi_0^2
-(\Pi_1+\frac{1}{2\pi})^2 \nonumber \\
&&+\Pi_0\Big [\Pi_2-\frac{1}{4\pi^2}
(v_{\uparrow\uparrow}-v_{\uparrow\downarrow})\Big ]\q^2 = 0.
\ee
The latter part (\ref{2}) yields the `out-of-phase' residues;
that is,
\be 
Res(K_{00},\omega({\q})) \sim
            \left( \begin{array}{cc} 1& \;  -1\\ -1& \; 1
                   \end{array}\; \right)\; .
\ee
Unlike the case of bilayer system\cite{lf3},
the residues for
the out-of-phase mode from (\ref{2}) yield  null contributions 
to the density response function 
after the summation over $\sigma$ in (\ref{pole}).

\begin{figure}
\inseps{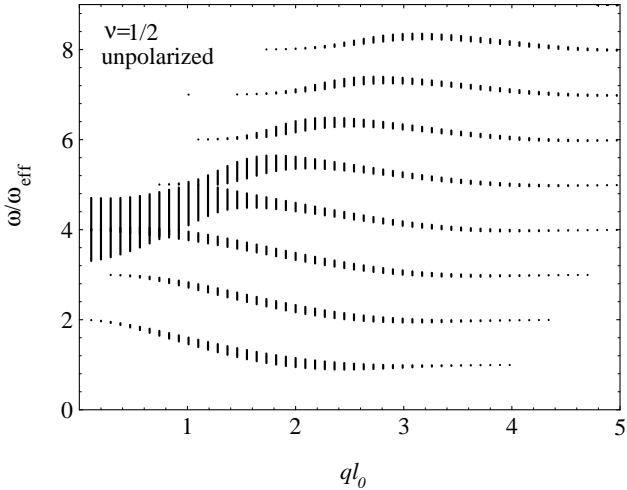}{0.5}
\caption{\label{fig4}
Collective excitation spectrum for both
the spin unpolarized state of $\nu=1/2$
with the short range potential $v(0)=4/5\hbar\omega_{\rm c}$.
The center of each stripe shows
the pole and width of striped band represents
$q^{-2}$ times the weights (residues) of the pole in $K_{00}$.}
\end{figure}

We find from 
(\ref{1}) with the use of (\ref{aa}) that there exists 
only one mode whose residue is proportional to $\q^2$
at the zero momentum frequency, $\o_{\rm c}=4\o_{\rm eff}$,
\be \label{w4}
\omega ({\q})= \Big [ { \omega}_{\rm c}^2 +
                         (6+\frac{6m_bv}{\pi}
                          {{\q}^2\over 2 {B_{\rm eff}}})\;
                         {\omega}_{\rm eff}^2   \Big ] ^{1\over2}
\ee
with 
\be \label{44}
Res(K_{00},\omega({\q})) = -{\q}^2 \omega_{\rm c} {\nu \over 8\pi}\;
            \left( \begin{array}{cc} 1& \;  1\\ 1& \; 1
                   \end{array}\; \right)\; ,
\ee
In Fig~\ref{fig4},
an almost flat minimum is
shown for the lowest energy mode.
As seen from (\ref{44}), the residue
for 
the mode of
zero momentum cyclotron frequency $4\omega_{\rm eff}$ is proportional
to $\q^2$ 
satisfying 
the Kohn's theorem \cite{kohn,sczhang}.  

There exists a large discrepancy
between our predicted energy gap and
the Zeeman energy; 
in the case of GaAs system 
the reallistic Zeeman gap is estimated to be $\sim 1.7{\rm K}$ for
$\nu=5/2$, and  $\sim 8.5{\rm K}$ for
$\nu=1/2$ with the choice of $B\sim 15$T for $\nu=1/2$ and 
$g$ $\sim 0.5$ for the $g$ factor,
while the computed energy gap is $\sim 40$K at the minimum of 
the lowest collective excitation mode.
To agree with the experiments\cite{willet}
which showed the compressible state at $\nu=1/2$ and
the incompressible state at $\nu=5/2$, 
the predicted energy gap of the singlet $\nu=1/2$ state should be 
substantially reduced.

\section{Conclusion}
In the present study, we introduced a spin-allowed $U(1)\times U(1)$
 Chern-Simons theory to extract all the known
 odd denominator filling factors 
 from a single Chern-Simons coupling constant (matrix)
 and derived 
 the formal expressions of spin-allowed
 electromagnetic polarization tensors
 corresponding to all the odd denominator filling factors.
Further we computed the collective excitation spectra
 for various  odd denominator filling factors
 in order to examine differences in the collective modes
between the different cases of
 the fully polarized states
 and the unpolarized or partially polarized states.

One of the salient features from the present CS theory is
that all possible 
odd denominator filling factors  corresponding to 
the spin-unpolarized, 
partially polarized,
and 
fully polarized states can be generated 
from 
a single CS coupling matrix.
By comparing
the collective excitation modes of $\nu=2/5$ and $3/7$ for the
cases of the spin-fully polarized and unpolarized or partially polarized states,
we find that both the unpolarized and partially polarized FQH states 
have much shallower minima for the lowest collective mode
compared to the fully polarized cases.
The Kohn's theorem \cite{kohn} was
satisfied for all the predicted collective excitation modes.
Judging from
the predicted shallower magnetoroton minima
with other than the spin fully polarized states,
we believe  that a study of finite-size system  through
the exact diagonalization procedure\cite{fano}
for the spin-unpolarized or partially polarized states
of odd denominator filling factors may also lead to shallower magnetoroton
minima than the fully polarized states.
In this paper, we did not allow the spin-flip and spin-wave mode.
Taking into account the spin-flip and spin-wave mode, 
the contributions
of exchange energy 
may deserve some attention utilizing
the present theory in the future.

\acknowledgments{
We greatly acknowledge helpful discussions with
Steven H. Simon
at Massachusettes Institute of Technology.
One (SHSS) of us acknowledges the generous supports of
the BSRI program (BSRI-95 and BSRI-96) of Korean Ministry of Education and the Center
for Molecular Sciences at Korea Advanced Institute of Science and Technology.
He is also grateful to Pohang University of Science and Technology (BSRI Program P96092) for  partial financial support.}

\widetext

\begin{references}
\bibitem{trick}
In order to compute the poles of
$K_{00}^{\sigma\sigma'}(\o, \q)$, we use the fact that the product of
$D^{\sigma\sigma'}(\omega,\q)$ by
$\Big [(\alpha^{\sigma \sigma'})^{-1} \Big ]^2$ cancels
its direct dependency on $\alpha^{\sigma \sigma'}$ for the
limiting case of $\varepsilon \rightarrow 0$.
\end{references}
\end{document}